\documentclass[twocolumn,aps,prl,tightenlines,amsmath,amssymb]{revtex4-1}
\usepackage{graphicx}
\usepackage{amssymb}
\usepackage{dcolumn}
\usepackage{amsmath}
\usepackage{bm}
\usepackage{colordvi}

\begin{document}

\title{Feasible optical weak measurements of complementary observables via a single Hamiltonian}
\author{Shengjun Wu$^1$ and Marek \.Zukowski$^{1,2}$}

\affiliation{$^1$ Hefei National Laboratory for Physical Sciences at Microscale and Department of Modern Physics, University of Science and Technology of China, Hefei, Anhui 230026, China\\
$^2$ Institute for Theoretical Physics and Astrophysics, University of Gdansk, PL-80952, Gdansk, Poland}

\date{\today}

\begin{abstract}

A general formalism for joint weak measurements of a pair of complementary observables is given. The  standard process of optical three-wave mixing in a nonlinear crystal (such as  in parametric down-conversion) is suitable for such tasks.
To obtain the weak value of a variable $A$ one performs weak measurements twice, with different initial states of the ``meter" field. This seems to be a drawback, but as a compensation we get for free the weak value of a complementary variable $B$. The scheme is tunable and versatile: one has access to a continuous set of possible weak measurements of pair of observables. The scheme increases signal-to-noise ratio with respect to the case without postselection.

\end{abstract}


\maketitle

The concept of weak measurements, introduced by Aharonov, Albert and Vaidman \cite{AAV}, gives us new insight and new tools to study quantum theory (for  experimental realizations see \cite{Ritchie,Pryde05,HK2008}).  It allows us to have a fresh look at counterintuitive quantum phenomena \cite{ABPRT02}, and investigate fundamental questions \cite{Wiseman07,Mir07,WJ08,Johansen}. It is also a practical tool for amplifying very tiny signals \cite{QSHE, Dixon, Starling, Resch, BS}, difficult to observe in the conventional schemes.
Very recently, with these ideas  a direct measurement of the quantum wavefunction  has been demonstrated
\cite{Lundeen2011}, a protocol for amplification of weak charge signals in the context of a solid-state setup has been presented \cite{ZRG2011},
and a scheme to amplify a cross-Kerr phase shift has been proposed \cite{FXS2011}.

Parametric down-conversion interaction (PDC),  see e.g. \cite{Pan2008rmp}, in the {\em stimulated} mode of operation,  seems to be a good candidate for a process via which one could realize weak measurements.
The coupling between the idler and signal beams is weak, and what is more important, it can be controlled by the strength of the pump beam, i.e., can be made as weak as we wish. However, the interaction Hamiltonian is at first glance totally unsuitable for this purpose, as it does not have the usual form of a coupling between a system observable ($\mathbf{A}$)  with an observable  ($p$) of the ``measuring device" (or rather measuring system), such as $g \mathbf{A} \otimes p$, where $g $ denotes the coupling strength. In this letter we propose the following.
One can treat the idler mode of the parametric process, denoted by $a_d$, as the ``device" or ``meter" subsystem, and the signal mode as the system to undergo a weak measurement  ($a_s$). If one as usual describes the (pulsed) pumping fields as classical, which is a standard, well working, approximation for narrow crystals and a not too strong pump, the parametric down conversion Hamiltonian can be put as
$\mathcal{H}= g f(t-t_0) (a_s a_d + a_s^\dagger a_d^\dagger)$, where $f$ represents a temporal overlap function of the pulse with the volume of the crystal (we can safely assume that it is very close to a  Dirac's delta $\delta(t-t_0)$), and $g$ is an effective coupling defined by the pump strength and geometry, second order non-linearity and crystal size (for a derivation and details see e.g. \cite{Pan2008rmp}). This form of interaction does not have the characteristic form required for weak measurements. However, if one puts  $\mathbf{B}=({a_s +a_s^\dagger})/{\sqrt{2}}$ and $\mathbf{A}=i({a_s -a_s^\dagger})/{ \sqrt{2}}$, and for the ``meter"  $q=({a_d + a_d^\dagger})/{\sqrt{2}}$ and
$p=i({a_d^\dagger} - a_d)/{\sqrt{2}}$, the Hamiltonian acquires a new form
\begin{equation}
 \mathcal{H}=g \delta(t-t_0) (\mathbf{A} \otimes p + \mathbf{B} \otimes q ),  \label{HAMILTONIAN}
\end{equation}
i.e., as the sum of two  Hamiltonians used for weak measurements, which apply to two {\em  complementary} situations, both for the system and the ``meter".  Notice that  $[q,p]=i,$ and $ [\mathbf{A},\mathbf{B}]=i$. We shall show below that, surprisingly,  a Hamiltonian of such a generic structure still allows weak measurements, and what is promising,  it is very versatile: one can also put it as $\mathcal{H}=g \delta(t-t_0) (\mathbf{A}' \otimes p' + \mathbf{B}' \otimes q' )$. The new operators are suitable pairs of quadratures
$\mathbf{B}'=({e^{-i\varphi}a_s +e^{i\varphi}a_s^\dagger})/\sqrt{2}$, $\mathbf{A}' =i ({e^{-i\varphi}a_s -e^{i\varphi}a_s^\dagger})/{ \sqrt{2}}$, $q'=({e^{i\varphi}a_d + e^{-i\varphi}a_d^\dagger})/{\sqrt{2}}$ and
$p'=i({e^{-i\varphi}a_d^\dagger} - e^{i\varphi}a_d)/{\sqrt{2}}$, for a given phase $\varphi$. We have the same structure for every $\varphi$!
We shall always put $\varphi=0$, however all that will be said below also equally pertains to other possible values.

Our claim is that a PDC-type Hamiltonian allows weak measurements of pairs of (complementary) observables.
Depending on the pairs of observables (defined by $\varphi$), one  has to prepare different initial states of the "measuring" subsystem, and different selection procedures. Still, the coupling remains the same, thus we have  in our labs  a ready, versatile,  tool for weak measurements. To obtain the weak value of a variable $\mathbf A$ one has to perform weak measurements twice (which is neither conceptually nor practically difficult). This seems to be a drawback, but as a compensation we get for free the weak value of the complementary variable $\mathbf B$.
Furthermore,  a weak measurement of a pair of observables may lead to a significant amplification of signal-to-noise ratio (SNR) with respect to the case without postselection (provided quantum noise, or shot noise, is dominant).  This is in sharp contrast with the case of weak measurements of
single observables. For such cases, once
quantum noise is dominant and the probability decrease due to
postselection is taken into account,   SNR and measurement sensitivity cannot be
effectively improved with respect to the case without postselection  \cite{zhu2011}.

Let us remark that a Hamiltonian like (\ref{HAMILTONIAN})  cannot generally lead to a decoherence of the  state of the system into an eigenbasis of neither $\mathbf{A}$ or $\mathbf{B}$. This may be the case, {\em e.g.},  for initial Gaussian states of the meter system. From the point of view of measurement theory  it does not provide an eigenselection, and thus is useless -- we stumble here on a new feature distinguishing weak and standard measurements.

Let us present a general description of
a weak measurement of two noncommuting observables  of a system, $\mathbf{A}$ and $\mathbf{B}$.  We assume that the interaction Hamiltonian of the system $s$ and a measuring. or ``meter", system $d$ is given, as above, by (\ref{HAMILTONIAN}).
In initial steps of the following  discussion neither the pair $(\mathbf{A}, \mathbf{B})$ nor the pair $(q, p)$ needs to be a complementary pair of observables.
We put $\hbar=1$ and assume that all factors in $g \mathbf{A} \otimes p$ and $g \mathbf{B} \otimes q$, especially $g$, are dimensionless.
We assume throughout that the adjustable constant $g$,  the effective strength of the system-meter interaction, is  small enough, so that  the weak interaction conditions \cite{WL2010}
$g \| \mathbf{A} \| \Delta p \ll 1$ and $g \| \mathbf{B} \| \Delta q \ll 1$ are satisfied. Thus, in perturbation expansion we  keep only terms of the lowest order in $g$.
Therefore,  the time-evolution  is given by
\begin{equation}
U = e^{-i g (\mathbf{A} \otimes p + \mathbf{B} \otimes q )} \approx 1- i g (\mathbf{A} \otimes p + \mathbf{B} \otimes q ) .
\end{equation}
Let $\left| \psi_i \right\rangle$ and $\left| \phi \right\rangle$ denote the initial states of the system and  of the meter, respectively.
After the interaction, the state of the compound  system  is given by
\begin{equation}
U \left| \psi_i \right\rangle \left| \phi \right\rangle  \approx
\left| \psi_i \right\rangle \left| \phi \right\rangle - i g (\mathbf{A} \otimes p + \mathbf{B} \otimes q ) \left| \psi_i \right\rangle \left| \phi \right\rangle .
\end{equation}
The subsequent postselection is defined by  projection onto the state of the system $\left| \psi_f \right\rangle$.
Upon a successful postselection, the final state (not normalized) of the measuring system
is given by
\begin{eqnarray}
\left| \phi ' \right\rangle &=&
\left\langle \psi_f \right| U \left| \psi_i \right\rangle \left| \phi \right\rangle  \nonumber \\
&\approx &
\left\langle \psi_f | \psi_i \right\rangle \left| \phi \right\rangle  - i g (\left\langle \psi_f \right| \mathbf{A} \left| \psi_i \right\rangle p + \left\langle \psi_f \right| \mathbf{B} \left| \psi_i \right\rangle q )  \left| \phi \right\rangle \nonumber \\
&=& \left\langle \psi_f | \psi_i \right\rangle \{ 1- i g  (\mathbf{A}_w p +\mathbf{B}_w q ) \} \left| \phi \right\rangle,
\end{eqnarray}
where $\mathbf{A}_w$ and $\mathbf{B}_w$ are the weak values defined according to
\begin{equation}
\mathbf{A}_w = \frac{\left\langle \psi_f \right| \mathbf{A}\left| \psi_i \right\rangle}{\left\langle \psi_f | \psi_i \right\rangle},
\label{weakvaluedefinitionold}
\end{equation}
and similarly for  $\mathbf{B}_w$.
The success probability of the postselection is given by
\begin{eqnarray}
P&=& \left\langle \phi ' | \phi ' \right\rangle   \nonumber \\
&\approx & \left| \left\langle \psi_f | \psi_i \right\rangle \right|^2
\left(1+ 2 g \left\langle p \right\rangle \textrm{Im} \mathbf{A}_w + 2 g \left\langle q \right\rangle \textrm{Im} \mathbf{B}_w \right).
\end{eqnarray}
 The average shift of the pointer variable $q$, conditional on preselection to $\left| \psi_i \right\rangle$ and postselection to $\left| \psi_f \right\rangle$, is given by
\begin{eqnarray}
\delta q &=& \frac{\left\langle \phi ' \right| q \left| \phi ' \right\rangle}{\left\langle \phi ' | \phi ' \right\rangle} - \left\langle \phi  \right| q \left| \phi \right\rangle \nonumber \\
&\approx & 2 g \textrm{Im} \mathbf{B}_w \textrm{var} q + g \textrm{Re} \mathbf{A}_w \left\langle - i [q, p]\right\rangle  \nonumber \\
&& + g \textrm{Im} \mathbf{A}_w \left\langle \{ q-\left\langle q \right\rangle, p- \left\langle p \right\rangle \}\right\rangle, \label{FORMULA1}
\end{eqnarray}
where $[t,s]$ and $\{t, s\}$ denote the commutator and anti-commutator of $t$ and $s$, respectively, $\left\langle s \right\rangle$ denotes the expectation value of $s$ in the initial state, $\textrm{var} q =\left\langle q^2 \right\rangle - \left\langle q \right\rangle ^2$ denotes the variance of $q$, or  $(\Delta q)^2$, in the initial state.
The average shift of the pointer variable $p$ is similarly given by
\begin{eqnarray}
\delta p &=& \frac{\left\langle \phi ' \right| p \left| \phi ' \right\rangle}{\left\langle \phi ' | \phi ' \right\rangle} - \left\langle \phi  \right| p \left| \phi \right\rangle \nonumber \\
&\approx & 2 g \textrm{Im} \mathbf{A}_w \textrm{var} p - g \textrm{Re} \mathbf{B}_w \left\langle - i [q, p]\right\rangle  \nonumber \\
&& + g \textrm{Im} \mathbf{B}_w \left\langle \{ q-\left\langle q \right\rangle, p- \left\langle p \right\rangle \}\right\rangle . \label{FORMULA2}
\end{eqnarray}
The approximations in calculating average pointer shifts are legitimate if the higher-order terms are negligible, therefore we further require that
$g \| \mathbf{A} \| \Delta p \ll | \left\langle \psi_f | \psi_i \right\rangle | $ and $g \| \mathbf{B} \| \Delta q \ll | \left\langle \psi_f | \psi_i \right\rangle | $. Obviously,  preselected and postselected states should not be orthogonal.

Let us now assume $q$ and $p$ are a pair of conjugate variables of the measuring device, i.e., $[q,p]=i$.  Therefore
\begin{eqnarray}
\delta q
&\approx & 2 g \textrm{Im} \mathbf{B}_w \textrm{var} q + g  \textrm{Re} \mathbf{A}_w   \nonumber \\
&& + g \textrm{Im} \mathbf{A}_w \left\langle \{ q-\left\langle q \right\rangle, p- \left\langle p \right\rangle \}\right\rangle   \label{shiftqpoition}\\
\delta p
&\approx & 2 g \textrm{Im} \mathbf{A}_w \textrm{var} p - g  \textrm{Re} \mathbf{B}_w   \nonumber \\
&& + g \textrm{Im} \mathbf{B}_w \left\langle \{ q-\left\langle q \right\rangle, p- \left\langle p \right\rangle \}\right\rangle . \label{shiftpmomentum}
\end{eqnarray}
For simplicity, we additionally assume that the meter is initially in a  Gaussian state
\begin{equation}
\phi_{\Delta q} (q) = \frac{1}{(\sqrt{2 \pi} \Delta q)^{1/2}} e^{-\frac{(q- q_0)^2}{4 (\Delta q )^2} + i p_0 q}.  \label{Gaussian}
\end{equation}
Thus, the anti-commutator terms in Eqs. (\ref{shiftqpoition},\ref{shiftpmomentum}) vanish. Therefore
\begin{eqnarray}
\delta q
&\approx & 2 g \textrm{Im} \mathbf{B}_w (\Delta q)^2 + g  \textrm{Re} \mathbf{A}_w     \label{shiftqpoitiongaussian}\\
\delta p
&\approx & 2 g \textrm{Im} \mathbf{A}_w (\Delta p)^2 - g  \textrm{Re} \mathbf{B}_w    \label{shiftpmomentumgaussian}.
\end{eqnarray}
Note that,  $\Delta p = 1 / (2 \Delta q)$ for the Gaussian state in (\ref{Gaussian}).

With the same interaction Hamiltonian but two different initial states $\phi_{\Delta q_1} (q)$ and  $\phi_{\Delta q_2} (q)$, we have
\begin{eqnarray}
\delta q _1&=& 2 g \textrm{Im} \mathbf{B}_w (\Delta q _1)^2 + g  \textrm{Re} \mathbf{A}_w   \\
\delta q _2&=& 2 g \textrm{Im} \mathbf{B}_w (\Delta q _2)^2 + g  \textrm{Re} \mathbf{A}_w   \\
\delta p _1&=&  2 g \textrm{Im} \mathbf{A}_w (\Delta p _1)^2 - g  \textrm{Re} \mathbf{B}_w    \\
\delta p _2&=&  2 g \textrm{Im} \mathbf{A}_w (\Delta p _2)^2 - g  \textrm{Re} \mathbf{B}_w  .
\end{eqnarray}
Thus,
\begin{eqnarray}
2 g \textrm{Im} \mathbf{A}_w &=& \frac{\delta p _1 -\delta p_2}{(\Delta p _1)^2 -(\Delta p _2)^2}  \\
2 g \textrm{Im} \mathbf{B}_w &=& \frac{\delta q _1 -\delta q_2}{(\Delta q _1)^2 -(\Delta q _2)^2}  \\
g  \textrm{Re} \mathbf{A}_w  &=&  \frac{(\Delta q _1)^2 \delta q _2 -(\Delta q _2)^2 \delta q _1}{(\Delta q _1)^2 -(\Delta q _2)^2}     \\
g  \textrm{Re} \mathbf{B}_w  &=&  \frac{(\Delta p _2)^2 \delta p _1 -(\Delta p _1)^2 \delta p _2}{(\Delta p _1)^2 -(\Delta p _2)^2}  .
\end{eqnarray}
Hence, with a pair of weak measurements, each with a different initial state of the meter, however both using the same Hamiltonian, or if one likes the same experimental setup,  one can determine  the real and imaginary parts of  the weak values of {\em a pair of non-commuting} observables.  The two initial states of the meter system must  differ in spreads of pointer variables $p$ and $q$.

In real experiments,  postselection may not be a projection onto a pure state, and the preselection might not be ideal. We extend the discussion to such cases in the Supplemental Material.

Thus far we have discussed the unexpected features of a PDC Hamiltonian. Let us now propose one of the possible
blueprints for an operational realization of such a scheme, employing standard tools of modern quantum optics, see  Figure 1.
With the laser pump beam on, the interaction within the crystal is described by (\ref{HAMILTONIAN}). Our ``system" and its initial state can be
a {\em weak} coherent state of light,  $\left| \alpha \right\rangle$. It is incident on a beam splitter which is slightly unbalanced. The output light from the  beam splitter is described by $\left| \psi_i \right\rangle =\left| \frac{\alpha}{\sqrt{2}}(1- \varepsilon)\right\rangle _{s'} \left| \frac{i \alpha}{\sqrt{2}}(1+ \varepsilon)\right\rangle _{s}, $ where $0<\varepsilon \ll 1$.
This will be our initial state of the system. The exit beam splitter in the system interferometer  is a symmetric 50-50 one.
The meter system, or ``pointer"  beam $d$,  is fed with a coherent or squeezed state. The proposed postselection of the system states corresponds to a detection of a single photon at the port 3s (and no photon at port 4s). The interferometer is tuned such that if the pump beam is off,  and thus there is no interaction within the crystal, and $\varepsilon=0$, only detector 4s registers light, i.e., 3s is a dark output in such a case. Formally this means that  we select states  $\left| \psi_f \right\rangle = \left| 1 \right\rangle_{3s}  \left| 0 \right\rangle_{4s} = \frac{1}{\sqrt{2}} ( \left| 1 \right\rangle_{s'} \left| 0 \right\rangle_s - i \left| 0 \right\rangle_{s'} \left| 1 \right\rangle_s )$.
In such a case the weak values of $\mathbf{A} = \frac{i}{\sqrt{2}}(a_s -a_s^\dagger)$ and $\mathbf{B}=\frac{1}{\sqrt{2}}(a_s +a_s^\dagger)$ are given by
\begin{eqnarray}
\mathbf{A}_w &\approx & - \frac{1}{2 \varepsilon \alpha} - \frac{\alpha}{2} \sim - \frac{1}{2 \varepsilon \alpha}; \label{WVofApure} \\
\mathbf{B}_w &\approx &  - \frac{i}{2 \varepsilon \alpha} + \frac{i \alpha}{2} \sim - \frac{i}{2 \varepsilon \alpha}. \label{WVofBpure}
\end{eqnarray}
Equations (\ref{shiftqpoitiongaussian}) and (\ref{shiftpmomentumgaussian}) give  the average shifts of the ``pointer"  beam in terms of $q$ and
 $p$. For the presented scheme they read:
\begin{eqnarray}
\delta q &=& - \frac{g}{\varepsilon} \frac{\textrm{Re} \alpha}{ |\alpha |^2} \left(  (\Delta q)^2 + \frac{1}{2}\right) ,  \label{qshiftlaser} \\
\delta p &=& \frac{g}{\varepsilon} \frac{\textrm{Im} \alpha}{ |\alpha |^2} \left(  (\Delta p)^2 + \frac{1}{2}\right) \label{pshiftlaser}.
\end{eqnarray}
The shifts $\delta q$ and $\delta p$ can be measured with a homodyne detection. Parameters  $\Delta q$ and $\Delta p$ depend on the initial state of the pointer beam.
As $\delta q$ and $\delta p$ are directly accessible, and $\varepsilon$ is also known and controllable, this scheme gives, for example, a direct method to measure the (small) coupling coefficient $g$.

Generally under suitable conditions one can achieve with weak measurements an amplification of signal and signal-to-noise rate.
If the final state of the system is not postselected, the average shifts of the pointer variables $q$ and $p$ are
$\delta q_0 \approx  g  \left\langle \mathbf{A}  \right\rangle_i     $ and
$\delta p_0 \approx  g  \left\langle \mathbf{B}  \right\rangle_i$,
where $\left\langle \Omega  \right\rangle_i \equiv \left\langle \psi_i | \Omega |\psi_i \right\rangle$ denotes the expectation value of
$\Omega$ in the initial state $\left| \psi_i \right\rangle$, which always lies inside the range of the eigenspectrum of $\Omega$.
By comparing these values with
(\ref{shiftqpoitiongaussian},\ref{shiftpmomentumgaussian}), we can calculate the signal amplification factors for our weak measurement scheme, defined as $K_q \equiv \frac{\delta q}{\delta q_0}$ and $K_p \equiv \frac{\delta p}{\delta p_0}$. As we shall show, each of them can
be much larger than 1. The reason for this is that the weak values could lie far outside of the eigenspectrum, and $\Delta q$ of the initial state is an adjustable parameter.

If no postselection is employed in reading of the shift of the pointer variable $q$, the signal-to-noise ratio (SNR) is given by $\mathcal{R}_q = \frac{\delta q_0}{\Delta q /\sqrt{N}}$, where $N$ denotes the number of repeated readouts, whereas the SNR for weak measurements is given by
$\mathcal{R}'_q = \frac{\delta q}{\Delta q /\sqrt{NP}}$, where $P\approx |\left\langle \psi_f | \psi_i \right\rangle |^2$ stands for the success probability of postselection.   To calculate $\mathcal{R}'_q$ we use the initial value of $\Delta q$. This is justifiable as in the lowest order of perturbation series, the standard deviation of $q$ of the pointer is not changed by the postselection. Therefore, the amplification factor of the SNR  $\mathcal{A}_q=\frac{\mathcal{R}'_q}{\mathcal{R}_q}$ is
\begin{equation}
\mathcal{A}_q= \frac{2 \textrm{Im} \mathbf{B}_w (\Delta q)^2 +  \textrm{Re} \mathbf{A}_w }{\left\langle \mathbf{A}  \right\rangle_i} |\left\langle \psi_f  | \psi_i \right\rangle | .  \label{amplifyq}
\end{equation}
In the case of the shift of $p$, due to the postselection in the weak measurement, the SNR is amplified by a factor
\begin{equation}
\mathcal{A}_p= \frac{2 \textrm{Im} \mathbf{A}_w (\Delta p)^2 -  \textrm{Re} \mathbf{B}_w}{\left\langle \mathbf{B}  \right\rangle_i } |\left\langle \psi_f  | \psi_i \right\rangle | . \label{amplifyp}
\end{equation}

\begin{figure}
\centering
\includegraphics[width=8cm]{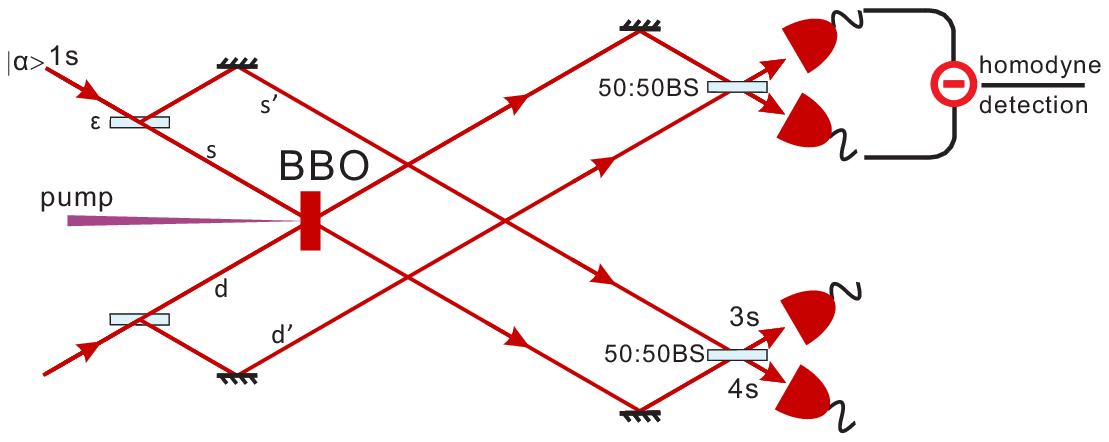}
\caption{(Color online) A sketch of a possible experimental setup. On a non-linear, say,  BBO crystal impinges a pump beam of a power tuned such that the three-wave mixing interaction leads effectively to the coupling Hamiltonian (\ref{HAMILTONIAN}) with a small enough $g$. Path $d$ represents the ``pointer" beam (device), whereas paths $s $ and $s'$ represent the measured system (beams are phase matched with the pump beams, which in turn is treated as classical). The exit beamsplitters are symmetric 50-50 ones. The upper input beamsplitter is slightly biased (see the text), whereas the lower one should be optimal  for the homodyne measurements. In the path $d$ leading to the BBO crystal one can put additional, active or passive,  optical elements modifying the properties of the ``pointer" beam. }
\label{fig1}
\end{figure}

To illustrate the above, and to show capabilities of the proposed experimental scheme, let us calculate in detail the amplification factors for the setup of fig. 1.
Without the postselection
the average shifts of the pointer position and the pointer momentum, as measured via a homodyne detection, are given by
$\delta q_0 \approx   - g \textrm{Re} \alpha $ and
$\delta p _0\approx   g \textrm{Im} \alpha$.
If the initial state of the pointer is a coherent one, we have $(\Delta q)^2 = (\Delta p)^2 =\frac{1}{2}$, whereas
$\delta q = - \frac{g}{\varepsilon} \frac{\textrm{Re} \alpha}{ |\alpha |^2}$
and
$\delta p =  \frac{g}{\varepsilon} \frac{\textrm{Im} \alpha}{ |\alpha |^2} $.
All this shows that in the weak measurements the measured shifts are amplified by a factor of $K =\frac{1}{\varepsilon |\alpha|^2}$. This amplification factor could be very large for a small  $\varepsilon |\alpha|^2$.

Let us move to the signal-to-noise ratios.
 Using (\ref{amplifyq}, \ref{amplifyp}) one can establish that the SNRs are amplified by a factor of
$\mathcal{A}= K \sqrt{P} = \frac{e^{-|\alpha|^2 /2}}{|\alpha|}$,
which approximately reads $\mathcal{A}\approx \frac{1}{| \alpha |}$ for $|\alpha| \ll 1$.
Therefore, by choosing a very small $\varepsilon$, we can have a very large signal amplification, while surprisingly by making the system beam very weak (i.e., a very small $|\alpha|$), both the shift and the SNR in the meter beam  can be amplified.
However, the amplification cannot be infinitely large. For the high-order terms of $g$ to be negligible, and thus all  the approximations legitimate, one must require that
$g \ll |\left\langle \psi_f | \psi_i \right\rangle | \approx \varepsilon |\alpha| $.
The overall success probability of postselection is $P \approx |\left\langle \psi_f | \psi_i \right\rangle |^2 \approx \varepsilon^2 |\alpha|^2$.
We use small $\alpha$ to achieve large SNR. This is especially useful in
cases in which the input beam is technically limited to very low number of
photons, or the quantum noise needs to be suppressed in order to register
observable pointer shifts.

For a squeezed state of the meter, the shifts $\delta q$ or $\delta p$ will be further amplified.  From  eqs. (\ref{qshiftlaser},\ref{pshiftlaser}), for  $\Delta q \gg 1/2$ we have $\delta q \approx - \frac{g}{\varepsilon} \frac{\textrm{Re} \alpha}{ |\alpha |^2}  (\Delta q)^2$ , and for  $\Delta p \gg 1/2$ we have $\delta p \approx  \frac{g}{\varepsilon} \frac{\textrm{Im} \alpha}{ |\alpha |^2}  (\Delta p)^2.$
For the aforementioned $\Delta q $ the signal $\delta q$ is amplified by a factor of
$K_q =\frac{(\Delta q )^2}{\varepsilon |\alpha|^2}$, and SNR by
$\mathcal{A}_q= K_q \sqrt{P} \approx \frac{(\Delta q )^2}{|\alpha|}$.

A typical thin BBO crystal, and a moderate pump power, lead to $g$ around $10^{-6}$.
If the meter beam is initially in a coherent state, for $\varepsilon \sim 10^{-2}$ and $\alpha \sim 10^{-2}$, we have an amplification of the signal by 6 orders of magnitude and an increase of SNR by 2 orders of magnitude.
If the meter beam is  prepared in a squeezed state, with $\Delta p$ or $\Delta q$ squeezed by a factor of 10,
a further amplification of  the signal and SNR by additional two orders of magnitude is possible.  Other sources of noise, e.g.,  systematic errors linked with alignment  in  the interferometers,
{\em etc}, may reduce the actual amplification factor. Note that the condition $g \ll |\left\langle \psi_f | \psi_i \right\rangle | \approx \varepsilon |\alpha| $ is satisfied for thin crystals with a not too strong pump. In such a  case, corrections to higher orders of $g$ is not necessary.

Finally, let us remark that a scheme for a weak measurement involving a PDC Hamiltonian was proposed in \cite{AP2007}.
However, in this case, the coupling was strong, the set-up worked in an optical parametric amplifier regime, and weak measurements were due to a weak coupling of the detector to the field in the idler mode. Thus the interaction (\ref{HAMILTONIAN}) was not directly responsible for the weak measurement, but rather used for preparation of initial states.  What we propose is absolutely different.
The PDC interaction Hamiltonian, in the weak coupling regime (a very small $g$), which at first glance looks unsuitable for an interaction, leads to weak measurement that has unexpected properties and is highly versatile.
It allows a quantum optical scheme for weak measurement of a pair of complementary observables. We can also choose which pair we are interested in.
The schemes described above are feasible, as the current methods of handling the PDC process are very well developed.
The general scheme for weak measurement of a pair of observables can also be implemented by other processes endowed with a  similar interaction Hamiltonian, such as the opto-mechanical interaction in the strong coupling regime, and the four-wave mixing process in cold atoms.

SW acknowledges support from NSFC (Grant No. 11075148),  the Fundamental Research Funds for the Central Universities, Chinese Academy of Sciences (CAS), and  National Fundamental Research Program.
MZ acknowledges EU Q-ESSENCE project, MNiSW (NCN) grant N202 208538 and a CAS visiting professorship.

\section*{Supplemental material: a discussion on imperfect pre- and postselection}

\subsection*{General case}
In real experiments, the postselection may not be a projection onto a pure state of the system, {e.g.}, detectors  not
sensitive to photon numbers can be modeled as a projection onto the subspace with at least one photon.
Suppose the initial state of the system is prepared in a general mixed state $\rho_s$ and the postselection is a general projection onto a subspace, denoted by $\Pi_f$.
With the evolution governed by (1) in the main text,  the state of the meter after the interaction and the postselection  is given by
$\rho_d ' = tr_s \left[ \Pi_f  U (\rho_s \otimes \rho_d ) U^{\dagger} \right]$, where $U$ is the time evolution operator due to the interaction, $\rho_d$ is the initial state of the measuring device, and the trace is taken over the Hilbert space of the system.  The average shift of a pointer variable $M$ conditional on the pre- and postselection of the system is given by
$\delta M = tr_d (M \rho_d ')/ tr_d(\rho_d ') -tr_d (M \rho_d)$.
After a detailed calculation, we get
 exactly the same formulas  (7-8)
as  for the case of pure  pre- and postselected  states in the main text. Of course,  the weak values ($\mathbf{A}_w$ and $\mathbf{B}_w$) for
general preselection $\rho_s$ and general postselection $\Pi_f$ are defined according to \cite{WL2010}
\begin{equation}
\left\langle \Omega \right\rangle _w \equiv \frac{tr(\Pi_f \Omega \rho_s)}{tr(\Pi_f \rho_s)},   \label{weakvaluedefinitionextended}
\end{equation}
where $\Omega= \mathbf{A}$ or $\mathbf{B}$.
This is a natural extension of the original definition of weak values.
A further    generalization  to positive
operator-valued  measurements (POVM) is possible, see \cite{DAJ10}, however we shall not address it here.

\subsection*{Our example}

Let us consider how the results in our proposed experiment can be affected, if the post-selected state is not a pure one.
Photons from the dark port 3s might be collected by a photon detector which cannot distinguish photon numbers, which is a quite typical situation.
The postselection is therefore better modeled as a projection onto a subspace with at least one photon, i.e., $\Pi_f =I_{3s} - \left| 0 \right\rangle _{3s} \left\langle 0 \right|$.
The pre-selected state can be written as $\left| \psi_i \right\rangle =\left| \frac{\alpha}{\sqrt{2}}(1- \varepsilon)\right\rangle _{s'} \left| \frac{i \alpha}{\sqrt{2}}(1+ \varepsilon)\right\rangle _{s}.$ At the output beamsplitter this leads to $\left| -\varepsilon \alpha \right\rangle _{3s} \left| i \alpha \right\rangle _{4s} $.
In terms of the variables of ports 3s and 4s, we have $$\mathbf{A} = \frac{i}{\sqrt{2}}(a_s -a_s^\dagger) =  \frac{1}{2} \left[ a_{3s} +a_{3s}^\dagger +i (a_{4s} - a_{4s}^\dagger) \right]$$ and $$\mathbf{B}=\frac{1}{\sqrt{2}}(a_s +a_s^\dagger) =  \frac{1}{2} \left[ -i (a_{3s} - a_{3s}^\dagger )+ a_{4s} + a_{4s}^\dagger \right].$$
Thus, from (\ref{weakvaluedefinitionextended}) we obtain the weak values
\begin{eqnarray}
\mathbf{A}_w &\approx & - \frac{1}{2 \varepsilon \alpha} - \textrm{Re} \alpha  \sim - \frac{1}{2 \varepsilon \alpha}; \\
\mathbf{B}_w &\approx & - \frac{i}{2 \varepsilon \alpha} - \textrm{Im} \alpha \sim - \frac{i}{2 \varepsilon \alpha}
\end{eqnarray}
which barely differ from Eqs. (22) and (23) in the main text, and thus can be approximated to the same values.
The correction is negligible for $\varepsilon \ll 1$, therefore the average shifts of $q$ and $p$ are still given by (24) and (25) in the main text.

For an even better modeling of the photon detection, one may take into account the cases in which the detector is not clicking, while there is one or more photons. Therefore, one can write $\Pi_f =\eta \left( I_{3s} - \left| 0 \right\rangle _{3s} \left\langle 0 \right| \right)$, where $\eta$ is the detector efficiency. Even with the detection efficiency taken into account, all the above results still hold since we only consider the data from a successful postselection.  The presence of an inefficiency only decreases the  probability of a successful postselection.

\end{document}